\documentclass[twocolumn,showpacs,amsmath,amssymb,pre]{revtex4}

\usepackage{graphicx}

\begin{document}

\title{Current behavior of a quantum Hamiltonian ratchet in resonance}

\author{Dario Poletti$^a$, Gabriel G. Carlo$^{b,a}$, Baowen Li$^{a,c,d}$}
\affiliation{$^a$ Department of Physics and Centre for Computational Science
and Engineering, National University of Singapore, Singapore 117542,
Republic of Singapore\\
$^b$ Departamento de F\'\i sica, Comisi\'on Nacional de Energ\'\i a At\'omica,
Avenida del Libertador 8250, 1429 Buenos Aires, Argentina\\
$^c$Laboratory of Modern Acoustics and Institute of Acoustics,
Nanjing University, 210093, P R China\\
 $^d$ NUS Graduate School for Integrative Sciences and Engineering,
117597, Republic of Singapore}
\date{\today}

\begin{abstract}
We investigate the ratchet current that appears in a kicked Hamiltonian system
when the period of the kicks corresponds to the regime of quantum resonance.
In the classical analogue, a spatial-temporal symmetry should be broken to
obtain a net directed current.
It was recently discovered that in quantum resonance the temporal symmetry can
be kept, and we prove that breaking the spatial symmetry is a necessary
condition to find this effect. Moreover,  we show numerically and analytically
how the direction of the motion is dramatically influenced by the
strength of the kicking potential and the value of the period.
By increasing the strength of the interaction this direction changes
periodically, providing us with a non-expected source of current reversals in
this quantum model.
These reversals depend on the kicking period also, though this behavior is
theoretically more difficult to analyze. Finally, we generalize the discussion
to the case of a non-uniform initial condition.
\end{abstract}


\maketitle

\section{Introduction}

The ratchet effect, a directed transport without any external net
force, has attracted an increasing interest since early studies by
Feynman et al.\cite{Feynman}. This phenomenon has a wide range of
possible applications in rectifiers, pumps, particle separation
devices, molecular switches, and transistors (see review articles
\cite{Hanggi,Reimann} and the references therein). It is also of
great interest in biology, since the working principles of
molecular motors can be conveniently explained in terms of ratchet
mechanisms \cite{Julicher}. Moreover, it is possible to
demonstrate quantum ratchet effects
\cite{Mennerat,Schiavoni,Meacher} by using cold atoms. As a
result, many different scenarios have been considered
\cite{Klafter,Schanz,Monteiro1,Monteiro2,Seba,Gong}.

At the classical level, directed transport in periodic systems
can be associated to a broken spatial-temporal symmetry
\cite{Flach}. We can obtain a net current by means of a
periodically kicked system, for instance. In this situation one
can break the spatial symmetry by using an asymmetric potential
and the temporal symmetry by introducing dissipation
\cite{gabriel} or an asymmetric kicking sequence \cite{Monteiro1}.
In these cases the quantum versions present the same symmetry
features as the classical counterparts, showing the corresponding
current.

Current reversals is one of the interesting ratchet features that has attracted
considerable interest \cite{RiemannPaper,Hanggi2,Mateos,Barbi,Dan}.
Besides the ones due to symmetries in the
potential, there are essentially two types of current reversals.
One corresponds to dissipative
chaotic systems and it is originated by bifurcations from chaotic
to periodic regimes, that is, by the transition from strange
attractors to simple ones \cite{Mateos}. The other \cite{RiemannPaper}
is explained by the fact that below certain temperatures, quantum
tunneling can cause a change in the direction of transport.

The system that we consider in this paper shows directed
transport associated to spatial asymmetry plus quantum resonance effects
rather than to an explicit spatio-temporal symmetry breaking. This kind
of systems was introduced in \cite{Lund} where the authors have found a
new mechanism for directed motion in quantum Hamiltonian systems.
In this kind of systems momentum grows indefinitely, i.e. it does
not stabilize around an asymptotic value (we could see this as a
"rectification of force").
In their work they have shown that even
if the system is time symmetric there can be transport in quantum
resonance. Quantum resonance (QR) is a pure quantum phenomenon
without a classical counterpart. In the well-known kicked rotor (KR)
system,
\begin{equation}
H=-\frac 1 2 \frac {\partial^2} {\partial \theta^2} +k \cos(\theta) \delta_T
\label{rotor}
\end{equation}
where $\delta_T=\sum_n\delta(t-nT)$, $T$ is the period of the
kick, $k$ the strength of the kick, and $\hbar$, the Planck
constant, and the moment of inertia of the rotator has been taken
equal to 1). Given a value of the kick strength $k$, special
resonant regimes of motion appear for periods with values $T=4\pi
\frac{r}{q}$, where the integers $r$ and $q$ are mutually prime.
Under these conditions the system regularly accumulates energy
which grows quadratically with the time and with $k$ \cite{Izrailev},
that is $\langle p^2\rangle \propto k^2t^2$, where $p$ stands for
the momentum operator (for simplicity we refer
to $\langle p \rangle$ as the momentum henceforth).

In the quantum KR at resonance (and for a symmetric or antisymmetric
initial condition) there is no growth of the momentum, i.e.,
it is always equal to its initial value (though it does grow like
$\langle p \rangle\propto kt$ if the initial condition is generic).
In order to find a net current for any initial condition, a different
potential has to be used. We prove that breaking any spatial
symmetry is a necessary condition. The potential that we study
corresponds to the double well-kicked rotor (dw-KR)\cite{Monteiro1}.
This potential has been experimentally realized in optical
lattices \cite{Weitz}.

In this model it is possible, for high resonances ($q>2$), to have
directed transport even for symmetric initial conditions. In addition,
this modified KR at resonance shows a new kind of current reversals
that (although being of quantum origin) are not due to tunneling
(like in \cite{RiemannPaper}). In fact, in our model the
momentum will evolve as $\langle p \rangle\propto g(k)t$ where $g(k)$
is a non monotonic function of $k$. This is quite different from that case in the
usual KR model. Moreover, we are able to give analytical
predictions for the current reversals following
a perturbative approach.

The paper is organized as follows. The main part of our paper, Section II, is
devoted to the numerical and analytical study of the behavior of the current,
including current reversals due to the variation of the kick strength and
period. We study this phenomenon analytically for small values of the
asymmetry. We also generalize this discussion to non-uniform initial
conditions. In Section III we show that breaking the spatial symmetry is a
necessary condition to find directed current at quantum resonance. In section
IV we present our conclusions.

\section{Directed current behavior}

The Hamiltonian of the system is given by:
\begin{equation}
H=-\frac 1 2 \frac {\partial^2} {\partial \theta^2} + k (\cos(\theta)+
a \cos(2\theta+\alpha))\delta_T.
\label{hamiltonian}
\end{equation}
where $\delta_T=\sum_n\delta(t-nT)$, $T$ is the period and $k$ is the
strength of the kick. This Hamiltonian can be used to study a gas of cold
atoms in an optical lattice. The difference between the dw-KR and the KR
is due to the parameters $a$, that is the relative strength of the second
harmonic and $\alpha$, which is a parameter that breaks the symmetry of the
kicking potential ($V(\theta)\neq V(2\pi-\theta)$ where
$V(\theta)= k (\cos(\theta)+ a \cos(2\theta+\alpha))$). The QR condition is
given by $T=4\pi r/q$, as in the KR case. Being $\hbar=1$, an effective
Planck's constant can be defined as $\hbar_{eff}=8\omega_RT=T$. It is
important to underline that there is no dissipation in our model. We study
the problem on a torus and the initial condition is $\phi_0=1/\sqrt{2\pi}$,
unless otherwise stated. This can be implemented on an optical lattice
with a wavefunction almost uniformly spread on many lattice sites. The
choice of this initial condition is due to the fact that it is symmetric.
This leads to the non typical behavior of this system compared with
the usual KR where for this scenario the momentum would
stay always at zero. Incidentally, it makes the analytical computations
much easier.
\begin{figure}[!h]
\includegraphics[width=\columnwidth]{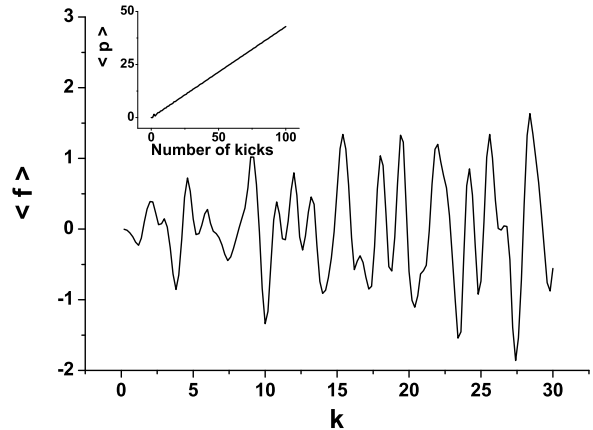}
\vspace{-0.6cm} \caption{The average effective force $<f>$ as a function
of the kick strength $k$, for $r/q=1/3$, $a=2$ and $\alpha=\pi/4$. In the
inset we show the momentum $\langle p\rangle$ versus the number of
kicks for $k=5$, $r/q=1/3$, $a=0.01$ and $\alpha=\pi/4$.}
\label{fig:overview}
\end{figure}

Firstly, we would like to make reference to the inset of
Fig.\ref{fig:overview}, where
we can see that for a given $k$, the momentum $\langle p\rangle$
grows linearly with time, that is with the number of kicks $N$
($\langle p\rangle\propto t\propto N$). For this numerical evaluation
we chose $T=4\pi\;1/3$, $k=5$, $a=0.01$ and $\alpha=\pi/4$.
Since $\langle p\rangle$ grows linearly with the number of kicks $N$
we focus our attention on a quantity defined as
$\langle f\rangle=\langle p\rangle/N$, which we call average
effective force. In Fig.\ref{fig:overview} we show the average
effective force versus the kick strength. We can see that there is a
general growth of $\langle f\rangle$ with $k$, but more interestingly, we
also see that $\langle f\rangle$ oscillates radically, going from positive
to negative values, i.e., there is current inversion.
To obtain this set of values we have used $a=2$, representative of the behavior of $\langle f\rangle$ for large values of $a$, and $\alpha=\pi/4$.
In order to see if this behavior can be detected in an experiment with
cold atoms, we have evaluated
$\langle p\rangle / \sqrt{\langle p^2\rangle-\langle p\rangle^2}$.
For this set of parameters,
we have found that it saturates to the value $0.18$ at about $15$ kicks
($N=15$), guaranteeing the feasibility of a future realization.

In the following we will show the origin of the current
inversions that occur for different values of the kick intensity or
the period. To do this we  do a perturbative study for small values of
the parameter $a$, which is the amplitude of the second harmonic.
In Ref. \cite{Izrailev} it has been shown that the one kick
evolution of an initial condition $\phi_0(\theta)$ is given by:
\begin{equation}
\phi_1(\theta)=\sum_{n=0}^{q-1}S_{0n}\phi_0(\theta+\frac{2\pi n}{q})
\label{eq:daizra}
\end{equation}
where $S_{0n}$ are given by $\beta_0(\theta)\gamma_n$ with
$\beta_{n}=\exp(-iV(\theta+\frac{2\pi n}{q}))$ and
$\gamma_n=\sum_{m=0}^{q-1}\exp(-i\frac{2\pi r m^2}{q}-i\frac{2\pi
m n}{q})$. Starting with a uniform initial condition we have:
\begin{equation}
\begin{array}{c}
\phi_1(\theta)=\beta_0\sum_{l_0=0}^{q-1}\gamma_{l_0}\phi_0(\theta+
\frac{2\pi l_0}{q})=
\frac{\beta_0}{\sqrt{2\pi}}\\
\\
\sqrt{2\pi}\phi_2(\theta)=\beta_0\sum_{l_1=0}^{q-1}\gamma_{l_1}\beta_{l_1}\\
\end{array}
\end{equation}
So, after a generic kick $N$ the wavefunction $\phi_{N+1}$ is given by:
\begin{equation}
\begin{array}{c}
\sqrt{2\pi}\phi_{N+1}(\theta)=\\
=\beta_0\sum_{l_1,...l_N=0}^{q-1}\gamma_{l_1}\beta_{(l_1+l_2+...l_N)}
\times\gamma_{l_2}\beta_{(l_2+...l_N)}...\times\gamma_{l_N}\beta_{l_N}\\
=\beta_0\sum_{A}\prod_{m,n=1}^{N}\gamma_{l_m}\beta_{l_n}
\end{array}
\end{equation}
where $A$ stands for the combinations given in the previous line of this
equation.

The momentum at time $N+1$ is given by
$\langle p_{N+1}\rangle=\int_0^{2\pi}\phi^*_{N+1}
(-i\frac{d}{d\theta}\phi_{N+1}) d\theta$ ($\hbar=1$) giving:
\begin{equation}
\begin{array}{c}
\langle p_{N+1}\rangle=\frac {-i} {2\pi}\int_0^{2\pi}
\big(\beta^*_0\sum_{A}\prod_{m,n=1}^{N}
\gamma^*_{l_m}\beta^*_{l_n}\big) \\
\big(\frac{d}{d\theta}  \big(\beta_0\sum_{A}\prod_{m,n=1}^{N}
\gamma_{l_m}\beta_{l_n}\big)\big)d\theta
\label{eq:mom}
\end{array}
\end{equation}
which can be rewritten as:
\begin{equation}
\begin{array}{c}
\langle p_{N+1}\rangle=\frac {-i} {2\pi}\int_0^{2\pi}
\big(\beta^*_0\sum_{A}\prod_{m,n=1}^{N}
\gamma^*_{l_m}\beta^*_{l_n}\big)\\ \big(\beta'_0\sum_{A}
\prod_{m,n=1}^{N}\gamma_{l_m}\beta_{l_n} +
\beta_0\sum_{A}\big(\frac{d}{d\theta}
\big(\prod_{m,n=1}^{N}\gamma_{l_m}\beta_{l_n}\big)\big)
\big)d\theta\\
\end{array}
\end{equation}
As we have shown in the inset of Fig.\ref{fig:overview} the
momentum grows linearly with $N$. The only part of the previous
expression that satisfies this condition is:
\begin{equation}
\begin{array}{cc}
\langle p_{N+1}\rangle\approx\frac {-i} {2\pi}\int_0^{2\pi}
\big(\beta^*_0\sum_{A}\prod_{m,n=1}^{N}
\gamma^*_{l_m}\beta^*_{l_n}\big)\\
\big(\beta_0\sum_{A}\big(\frac{d}{d\theta}\big(\prod_{m,n=1}^{N}
\gamma_{l_m}\beta_{l_n}\big)\big)\big)d\theta
\end{array}
\label{eq:momint}
\end{equation}
because we have the derivative of a product of N terms.\\
Hence, we can approximate Eq.(\ref{eq:mom}) by:
\begin{equation}
\begin{array}{c}
\langle p_{N+1}\rangle\approx N \frac {-i} {2\pi}\Big(\int_0^{2\pi}
\big(\sum_{m\neq n}\gamma^*_{m}\beta^*_{m}
\gamma_{n}\beta'_{n}\big)d\theta\;+\\
+\;\int_0^{2\pi}
\big(\sum_{l,m,n,s}\gamma^*_{l}\gamma^*_{m}\gamma_{n}\gamma_{s}\beta^*_{l+m}
\beta^*_{m}
\left(\beta_{n+s}\beta_{s}\right)'\big)d\theta\;+\;...\Big)\\
\label{eq:idea}
\end{array}
\end{equation}
Here we make a further approximation, we drop all the integrals with
the exception of the first one. This is justified by the fact that their
integrands are highly oscillating functions of $\theta$ compared to
the first one, so their contribution to the final result is negligible.

In order to simplify the analytic treatment without losing the essential
features of $\langle p\rangle$ that we want to describe,
we focus on the usual KR
perturbed by a second harmonic, i.e., we take $k\sim O(1)$ and
$a\ll 1$ . In this case, as long as $ka\ll 1$ we can approximate
$\beta_m\simeq\tilde{\beta}_m= \exp(-ik\cos(\theta+2\pi
m/q))(1+ika\cos(2\theta+4\pi m/q+\alpha))$. Then,
Eq.(\ref{eq:idea}) becomes:

\begin{equation}
\langle p_{N+1}\rangle\approx N \frac {-i} {2\pi}\int_0^{2\pi}
\big(\sum_{m,n}\gamma^*_{m}
\gamma_{n}V'(\theta+2\pi n/q)\tilde{\beta}^*_{m}\tilde{\beta}_{n}
\big)d\theta\\
\label{eq:momsmalla}
\end{equation}
These terms are all integrable and after a few computations we can
write the momentum as $\langle p_{N+1}\rangle\approx \sum_{m,n}
L_{m,n}$ where $L_{m,n}$ is given by:

\begin{equation}
\begin{array}{c}
L_{m,n}=k\;a\;N\gamma^*_m\gamma_n \\
\Big( \sin\big(\frac{2\pi n}{q} - \omega_{m,n}\big)\times\\
\times\left[  \cos\big(\frac{4\pi n}{q} + \alpha - 2\omega_{m,n}\big) -
\cos\big(\frac{4\pi m}{q} + \alpha - 2\omega_{m,n}\big)  \right]\times\\
\times\big(k\;J_1(\omega_{m,n}k)-2\;J_2(\omega_{m,n}k)/\omega_{m,n}\big)+\\
-\cos\big(\frac{2\pi n}{q} - \omega_{m,n}\big)\times\\
\times\left[  \sin\big(\frac{4\pi n}{q} + \alpha - 2\omega_{m,n}\big) -
\sin\big(\frac{4\pi m}{q} + \alpha - 2\omega_{m,n}\big)  \right]\times\\
\times2\;J_2(\omega_{m,n}k)/\omega_{m,n}+\\
-2\sin\big(\frac{4\pi n}{q} + \alpha - 2\omega_{m,n}\big)J_2(\Omega_{m,n}k)
\Big)
\label{eq:momsol}
\end{array}
\end{equation}

where $\Omega_{m,n}=\sqrt{\mu^2_{m,n}+\nu^2_{m,n}}$ and
$\tan\omega_{m,n}=\big(\frac{\nu_{m,n}}{\mu_{m,n}}\big)$.
Here we have used $\mu_{m,n}=\left[\cos(2\pi n/q)-\cos(2\pi m/q)\right]$ and
$\nu_{m,n}=\left[\sin(2\pi n/q)-\sin(2\pi m/q)\right]$. The momentum
$\langle p\rangle$ depends on the period parameter $r$ through
the coefficients $\gamma_m$.
We now see that it is not surprising to find
current inversion. In fact, we expect a different sign of
the average effective force for different values
of $k$, because the Bessel functions oscillate.

From the properties of the Bessel functions we can understand the
behavior for small and large $k$. For the case of $k\gg 1$,  we must guarantee
that $ka\ll 1$. In general we have
that for $x\ll 1$, $J_{\alpha}(x)\simeq \frac 1 {\Gamma(1+\alpha)}
\big(\frac x 2 \big)^{\alpha}$ ($\Gamma$ is the function which generalizes
the factorial to non-integer numbers) and for $x \gg 1$ we have that
$J_{\alpha}(x)\simeq \sqrt{\big(\frac 2 {\pi x}\big)}\cos(x-\alpha/4-\pi/2)$.
For higher values of the period parameter $q$ ($T=4\pi\; r/q$),
the situation is  more complex because
we have many different $\Omega_{m,n}$ and so there is a superposition of
many terms like $\cos(\Omega_{m,n}k-\alpha\pi/4 -\pi/2)$. This is much
more difficult to study analytically.

In general, for large values of $k$, we can approximate the
behavior by
\begin{equation}
\begin{array}{c}
\langle p_{N+1}\rangle\approx \sum_{m,n} \big[A\big(\frac{k^3}{\Omega_{m,n}}\big)^{1/2}
\cos(\Omega_{m,n}k-\frac{3\pi}{4})+ \\
B\big(\frac{k}{\Omega_{m,n}}\big)^{1/2}
\cos(\Omega_{m,n}k-\frac{5\pi}{4})\big],
\end{array}
\end{equation}
where $A$ and $B$ are two constants that can be obtained from Eq.(\ref{eq:momsol}).
We can think of this expression as divided into two parts. The first part
dominates for large values of $k$ when it is far enough from its zeros.
Even if we consider only the first part, the sum of many cosine
functions with different periods gives a very fluctuating behavior.

To confirm the above analysis, we compute the particular cases
$r/q=1/3$ and $r/q=1/5$ and compare them with the numerical
solution obtained from the evolution of the wavefunction. In the
case of $r/q=1/3$, $\Omega_{m,n}$ takes only two values, either $0$ or
$\sqrt{3}$ and so we only have terms like $J_{\eta}(\sqrt{3}k)$
where $\eta=1,2$.

After some analytical computations we can see that the average effective force is
given by:
\begin{equation}
\begin{array}{c}
\langle f\rangle=k\;a\;\sin(\alpha)\\
\left[\left(\frac 1 {\sqrt{3}}-1\right)k\; J_1\left(\sqrt{3}k\right)+
\frac 2 3 \left(1+\sqrt{3}\right)\;J_2\left(\sqrt{3}k\right)\right]
\label{eq:approxforce}
\end{array}
\end{equation}

\begin{figure}[!h]
\includegraphics[width=\columnwidth]{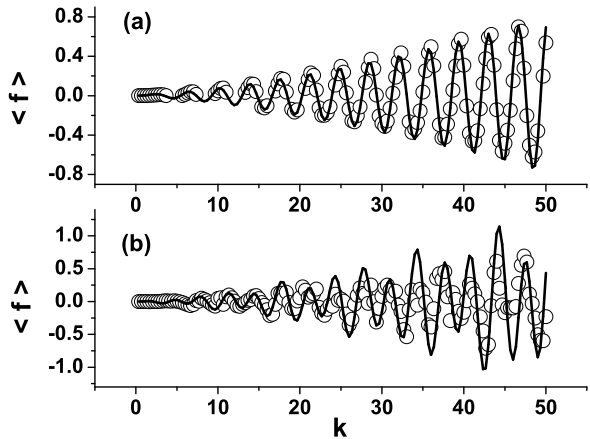}
\vspace{-0.6cm} \caption{Average effective force $\langle f\rangle$
versus $k$ for $a=0.01$ and $\alpha=\pi/3$. Results from the numerical
evolution of the wavefunction (circles) and
the analytical approximation (solid line) are compared. In (a) $T=4\pi\;1/3$
while in (b) $T=4\pi\;1/5$. We can see the oscillations showing current
reversals.}
\label{fig:speed3}
\end{figure}

The case for $r/q=1/3$ is shown in Fig.\ref{fig:speed3}(a) where the
numerical (circles) and analytical (solid line) results are compared
and show good agreement. A different period implies different values of
$\Omega_{ij}$ and so different values of the period of oscillation for
the same value of $k$. We should expect very different behavior
for different $r/q$. This is clarified in Fig.\ref{fig:speed3}(b)
where we plot the numerical (circles) and analytical (solid line)
results showing the average effective force versus $k$ for $r/q=1/5$.
In this case, differences are due to the cut-off of oscillating terms.

It is interesting to notice one important difference in this model
compared to the usual quantum KR model. In our model the average effective
force does not grow linearly with the intensity of the kick. Instead, it
oscillates periodically. Even the peaks do not show this linear
behavior with respect
to $k$. There are two components contributing to the average effective
force. One of them causes the peak to grow with a power $1/2$
($\langle f\rangle \propto k^{1/2}$, i.e. $\langle p\rangle\propto k^{1/2}t$) and the other makes
it to grow with a power $3/2$ ($\langle f\rangle\propto k^{3/2}$).

Reversals due to changes in the value of the kick strength
$k$ have been found previously \cite{Hutchings}, but in a completely
different context.
Here we find reversals due to changes in k and also in
the value of the period, i.e. in the effective Planck
constant. Since in our case the nature of the current is different
from the one in \cite{Hutchings}, these inversions
are also of a different and novel character \cite{Gong}.
An insightful model of QRs in the quantum KR can be found in
\cite{Wimberger}, and we think that a theoretical explanation for
this phenomenon could be investigated following those lines.

\begin{figure}[!h]
\includegraphics[width=\columnwidth]{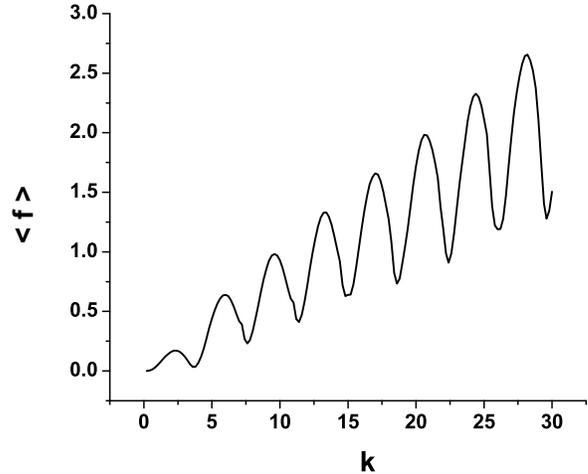}
\vspace{-0.6cm} \caption{Average effective force $\langle f\rangle$
versus $k$ for $T=4\pi\;1/3$, $a=0.01$ and $\alpha=\pi/3$. The initial
condition is $\phi_0(\theta)=\eta\cos(\cos(\theta)+\sin(2\theta))$,
where $\eta$ is the normalization constant. } \label{fig:nonuni}
\end{figure}

In the generic case, the initial condition will not be symmetric,
a situation which is most likely to happen in real experimental
conditions. As already discussed, for an asymmetric initial condition
and a symmetric kick, quantum resonance induces directed motion
and the momentum will increase linearly with the time and with
the strength of the kick. On the other hand, with a non
symmetric kick in quantum resonance, we will see the
oscillating behavior that we have found previously, superimposed in this
case to a linear growth due to the asymmetry
of the initial condition. This is confirmed by the numerical
results shown in Fig. \ref{fig:nonuni}. We have considered $T=4\pi\;1/3$,
$a=0.01$, $\alpha=\pi/3$ and the initial condition
$\phi_0(\theta)=\eta\cos(\cos(\theta)+\sin(2\theta))$,
where $\eta$ is the normalization constant.

\section{Necessary condition for directed transport in quantum resonance}

In this section, we shall show that in the absence of a net
force, either an asymmetry in the initial condition or in the
kicking potential is necessary to have directed transport. It should be stressed that 
these are necessary but not sufficient
conditions, because it is possible to have zero transport also
when at least one of these two asymmetries is present. Representing a gas of cold
atoms with the wavefunction $\phi$, the momentum is given by
$\langle p\rangle= -i\hbar\int_0^{2\pi}\phi^*\big(\frac d
{d\theta} \phi\big) d\theta$. Let's take again a generic kicked
system with a Hamiltonian given by:

\begin{equation}
H=-\frac 1 2 \frac {\partial^2} {\partial \theta^2} +
V(\theta)\delta_T
\label{genham}
\end{equation}
where $\delta_T=\sum_n\delta(t-nT)$, $T$ is the period
of the kick and $V$ is the shape of the kicking potential.
We assume that the system is at quantum resonance
($T=4\pi \frac r q$ with $r$ and $q$ relative
prime numbers). Given the wavefunction after $N$ kicks ($\phi_N$),
we can obtain the wavefunction after $N+1$ kicks $\phi{_N+1}$ by
\cite{Izrailev}:
\begin{equation}
\phi_{N+1}(\theta)=\beta_0\sum_n\gamma_n\phi_N\Big(\theta+2\pi \frac n q\Big)
\label{eq:qrevo}
\end{equation}
where $\beta_{n}=\exp(-iV(\theta+\frac{2\pi n}{q}))$ and $\gamma_n=
\sum_{m=0}^{q-1}\exp(-i\frac{2\pi r m^2}{q}-i\frac{2\pi m n}{q})$.
Using the definition of $\gamma$ one can verify that
\begin{equation}
\gamma_n=\gamma_{n+q}=\gamma_{q-n}.
\end{equation}
Let's study the case with a symmetric initial
condition ($\phi_N(2\pi-\theta)=\phi_N(\theta)$) and a
symmetric kicking potential ($V(2\pi-\theta)=V(\theta)$).
In this case we can see that $\beta_n(2\pi-\theta)=\beta_{q-n}(\theta)$ and then
\begin{equation}
\begin{array}{c}
\phi_{N+1}(2\pi-\theta)=
\beta_0(2\pi-\theta)\sum_n\gamma_n\phi_N\Big(2\pi-\theta+2\pi \frac n q\Big)\\
\phi_{N+1}(2\pi-\theta)=\beta_0(\theta)
\sum_n\gamma_{q-n}\phi_N\Big(\theta+2\pi \frac {q-n} q\Big).\\
\end{array}
\end{equation}
This shows that given a symmetric initial condition and a
symmetric kicking potential, the wavefunction will always be
symmetric and therefore the momentum will always be zero. This
result is readily extended to anti-symmetric initial conditions.

We now show that if the
initial condition has a shape like $\phi_N(\theta)=e^{iL\theta}f_N(\theta)$
where $f_N(\theta)=f_N(2\pi-\theta)$ is even, the kick will not
change the value of the momentum which is $\langle p\rangle=L$.
For example, an eigenstate of the momentum
$\phi_0=\frac {e^{iL\theta}} {\sqrt{2\pi}}$ will not change its
direction or speed. Using Eq.(\ref{eq:qrevo}) it is easy to show that:
\begin{equation}
\begin{array}{c}
\phi_{N+1}(\theta)=
\beta_0(\theta)\sum_n\gamma_ne^{iL(\theta+2\pi \frac n q)}f_N
\Big(\theta+2\pi \frac n q\Big)=\\
=\beta_0(\theta)e^{iL\theta}\sum_n\gamma_ne^{iL(2\pi \frac n q)}f_N
\Big(\theta+2\pi \frac n q\Big)\\
\end{array}
\end{equation}
where we have factorized out the $e^{iL\theta}$ which contributes to
the  momentum and now the factors
$\gamma_n$ are multiplied by a phase $e^{iL(2\pi \frac n q)}$.
The symmetry of $\phi_{N+1}(\theta)$ is given by:
\begin{equation}
\begin{array}{c}
\phi_{N+1}(2\pi-\theta)=\\
=\beta_0(2\pi-\theta)\sum_n\gamma_ne^{iL(2\pi-\theta+2\pi \frac n q)}
f_N\Big(2\pi-\theta+2\pi \frac n q\Big)=\\
=\beta_0(\theta)e^{iL(2\pi-\theta)}\sum_n\gamma_{q-n}e^{iL(2\pi \frac n q)}
f_N\Big(\theta+2\pi \frac {q-n} q\Big)=\\
=\beta_0(\theta)e^{iL(2\pi-\theta)}\sum_n\gamma_{q-n}e^{iL(2\pi \frac {q+n} q)}
f_N\Big(\theta+2\pi \frac {q-n} q\Big)
\end{array}
\end{equation}
Using the fact that $\gamma_ne^{iL(2\pi \frac n q)}=e^{-iL^2(2\pi \frac p q)}
\gamma_{n+2L}=e^{-iL^2(2\pi \frac p q)}\gamma_{q-n+2L}=e^{-iL^2(2\pi \frac p q)}
\gamma_{q-n-2L}$ it is possible to show that:
\begin{equation}
\gamma_ne^{iL(2\pi \frac n q)}=\gamma_{q-n}e^{iL(2\pi \frac {q-n} q)}
\label{eq:propgam}
\end{equation}
From Eq.(\ref{eq:propgam}) it is evident that after any kick
$\phi_N(\theta)$ can be decomposed as $\phi_N(\theta)=e^{iL\theta}f_N(\theta)$
and this proves that the momentum will not change.

It is well known that with an asymmetric initial condition we can
have transport. In fact, for the simple case $T=4\pi$,
we have that $\phi_N=e^{-iNV(\theta)}\phi_0$.
If $\phi_0$ is not symmetric, even if we have
zero net force the momentum grows linearly with
the number of kicks.
In \cite{Lund}, it has been found
that with a non-symmetric kick, transport can be induced by high $q$
resonances. This means that it is necessary to at least break
the spatial symmetry to have
transport. To do this we can either start with a non-symmetric initial
condition, use a non-symmetric kick or a combination of the above situations.

\section{Conclusions}

We have analyzed in detail the phenomenon of directed transport
for a Hamiltonian system in quantum resonance. We have found that
directed transport can be in different directions depending on the
intensity of the kick or on its period. This phenomenon was
unexpected and presumably due to how the intensity of the kick and
the value of the period affect the gradient of the phase of the
wavefunction. Very interestingly, the direction of the motion
changes periodically with the intensity of the kick. Also
remarkable is the fact that the momentum at the peaks does not
increase linearly with the intensity of the kick, but it follows
one term that goes as a square root and another that goes as a
power law with exponent $3/2$.
We have also shown that for an asymmetric initial
condition the periodic effect due to quantum resonance is superimposed
to the expected drift due to the asymmetry of the initial
condition. Finally, we have found that even though an explicit
time symmetry breaking is not needed, breaking the spatial
symmetry is a necessary condition to have directed transport.
By numerical simulations we
have also found out that this effect is very sensitive on the
initial condition. To see clearly these effects experimentally a
non-interacting BEC could be used because the initial wavefunction is
less-spread in the momentum space than for a gas of cold atoms.

\section{acknowledgments}
We would like to thank  G. Casati,  G. Benenti and  W. Wang for
fruitful discussions. G.G.C. gratefully acknowledges support by
Conicet (Argentina). This work is also supported in part by a FRG
grant of NUS and the DSTA under Project Agreement POD0410553.

\end{document}